\documentstyle[multicol,epsf,aps]{revtex}
\def\ruleleft{\vspace{-2.5\baselineskip}\begin{multicols}{2}\ \linebreak\vspace{-\baselineskip}\hrulefill\raisebox
{0.84mm}{$\!\rfloor$}\[\]\end{multicols}\vspace{-1.5\baselineskip}}
\def\ruleright{\vspace{-1.5\baselineskip}\begin{multicols}{2}\ \linebreak\raisebox
{-2.45mm}{$\lceil\!$}\hrulefill\end{multicols}\vspace{-\baselineskip}}
 
\newlength{\normalarraycolsep}
\newlength{\normaltabcolsep}
\newcommand{\bm}[1]{\mbox{\boldmath $#1$}}
\newcommand{\bms}[1]{\mbox{\scriptsize \boldmath $#1$}}
\newcommand{\rb}{\bm{r}}
\newcommand{\Eb}{\bm{E}}
\newcommand{\fracd}[2]{\frac{\displaystyle#1}{\displaystyle#2}}
\newcommand{\be}{\begin{equation}}
\newcommand{\ee}{\ \ \ \end{equation}}
\newcommand{\bea}{\setlength{\arraycolsep}{0.4\normalarraycolsep}
                  \begin{eqnarray}} 
\newcommand{\eea}{\ \ \ \end{eqnarray}\setlength{\arraycolsep}
                  {\normalarraycolsep}}
\newcommand{\bean}{\setlength{\arraycolsep}{0.4\normalarraycolsep} 
                  \begin{eqnarray*}}
\newcommand{\eean}{\ \ \ \end{eqnarray*}\setlength{\arraycolsep}
                  {\normalarraycolsep}}

\begin{document}
\title{\bf Donnan equilibrium and the osmotic pressure of charged 
colloidal lattices}
\author{\bf M\'ario N. Tamashiro, Yan Levin and Marcia C. Barbosa}
\address{\it Instituto de F\'{\i}sica,
Universidade Federal do Rio Grande do Sul\\
 Caixa Postal 15051, 91501-970 Porto Alegre (RS), Brazil \\
{\small mtamash@if.ufrgs.br, levin@if.ufrgs.br, barbosa@if.ufrgs.br}}
\maketitle
\begin{abstract}
We consider a system composed of a monodisperse charge-stabilized
colloidal suspension in the presence of monovalent salt,
separated from the pure electrolyte by a semipermeable 
membrane, which allows the crossing of solvent, counterions, and 
salt particles, but prevents the passage of polyions.
The colloidal suspension, that is in a crystalline phase, is 
considered using a spherical Wigner-Seitz cell.
After the Donnan equilibrium is achieved, there will
be a difference in pressure between the two sides of the membrane.
 Using the  functional
density theory, we obtained the expression for the osmotic
pressure as a function of the  concentration of added salt, 
the colloidal volume fraction, and the size and charge of the 
colloidal particles. The results are compared with  the experimental 
measurements for ordered polystyrene lattices of two different 
particle sizes over a range of ionic strengths and colloidal 
volume fractions. 
\end{abstract}
\medskip

\centerline{{\bf PACS numbers:} 82.70.Dd; 36.20.$-$r; 64.60.Cn}

\begin{multicols}{2}
\section{Introduction}

In recent years, colloidal particles have received an increased attention
because they constitute an interesting system from practical, experimental,
and theoretical point of view.
Large molecules immersed in a solution  are important in various
systems, from biological to industrial. Due to the large size of the 
particles,
a rich variety of experiments can be easily performed.  
Optical measurements show that some suspensions, such as the opals 
\cite{Sa64} and
the viruses \cite{St35}, form regular lattices that can, in principle, 
exhibit melting and
structural phase transitions \cite{Re97,Wi74}. The elastic rigidity of
these ordered structures leads to unusual viscoelastic properties. 
The suspension responds to small-amplitude deformations as a linear
viscoelastic solid, allowing for the propagation of the 
low-frequency shear waves \cite{Ru81,Be81}.

 From the purely theoretical point of view, colloids are quite fascinating 
materials
to study. Although thermodynamically identical to the usual atomic 
fluids and solids, colloids have an additional advantage in that
the range of the interactions between the colloidal particles can be 
``manually''
controlled by exploiting their interactions with a surrounding solvent, as 
well as by actually synthesizing particles which behave in a desired fashion.
In general, the interactions between the colloids are dominated by a van der Waals
force resulting from the quantum fluctuations of the electron charge density
on the surface of a colloidal particle. This attractive interaction can lead to
an aggregation of macromolecules and to their precipitation in a gravitational 
field. In many practical applications, such as design of water-soluble paints,
it is essential to device a mechanism which would stabilize the colloidal suspension
against precipitation. One such mechanism is to synthesize colloidal particles with 
some acidic groups on their surfaces, which will be ionized upon contact with water.
The sufficiently strong electrostatic repulsion between the equally charged 
macromolecules will prevent the formation of clusters and stabilize the suspension
against the precipitation. It is the aim of this paper to try and shed some light
on the behavior of charge-stabilized colloids.

When the volume fraction is not too small, 
the charged polyions form an ordered structure (bcc or fcc)
\cite{Gr91b,Al84,Ha96}. 
The description of the suspension in this case
becomes significantly more simple than that  of a
disordered structure \cite{Le97,Ta92,Cr94}, since
one can take advantage
of the translational symmetry of the lattice.
Thus each polyion, with its counterions, is
inclosed in a Wigner-Seitz {\it (WS)}\/ cell 
\cite{Gr91b,Al84,Ha96}.
The thermodynamics of the system is then fully determined
by the behavior inside one cell.
In this paper we will present a  simple theory
to calculate the Donnan equilibrium properties of a charged 
colloidal lattice. The central problem will be to obtain the 
osmotic pressure of a colloidal crystal in equilibrium 
with a reservoir of salt. The experimental setup is described
in Ref. \cite{Vi81}. It consists of an osmometer, which is made 
of two cells separated by a semipermeable membrane, which allows for
the crossing of solvent, counterions, and the microions of salt, but 
prevents the passage of polyions. This, in turn, leads to an establishment 
of a membrane potential and an imbalance in the number of microions
on  the two sides of the membrane, resulting in an osmotic 
pressure measured by two capillaries attached to the chambers containing
the colloidal suspension and the pure salt 
solution. This osmotic pressure, measured in
mm of H$_2$O, will be compared to that found on the basis of our
theoretical calculations.

\section{Osmotic pressure}

Since the  colloidal suspension is organized in a periodic structure, 
it is sufficient to consider just one isolated
polyion in a salt solution inside an appropriate 
{\it WS}\/ cell. 
To simplify the problem, a further approximation is to
replace the polyhedral {\it WS}\/  cell by a sphere of the
same volume $V=\fracd{4\pi}{3}R^3$, as represented in Fig.~1.
Previous calculations, which 
compare explicitly results derived from the spherical-cell 
model with those that follow from a periodic system of cubic 
symmetry, show that this approximation provides a good 
description of the multibody system at low-volume 
fractions \cite{Gr91b}.
\begin{figure}[ht]
\begin{center}
\leavevmode
\epsfxsize=0.35\textwidth
\epsfbox[50 30 595 560]{"
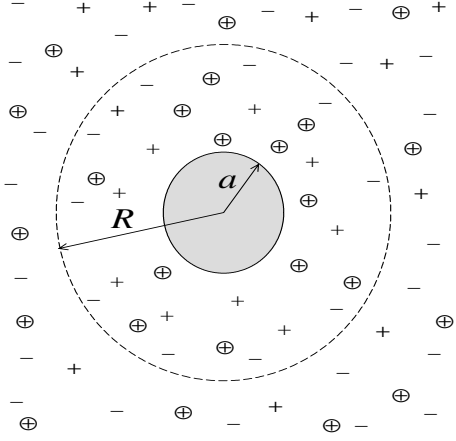"}
\end{center}
\begin{minipage}{0.48\textwidth}
\caption{Geometry of the spherical Wigner-Seitz cell with radius $R$.
The mesoscopic polyion (with total charge $-Ze$ distributed uniformly 
on its surface) is represented by the shaded sphere of radius $a$. 
The semipermeable membrane located at $|\rb|=R$, represented by the 
dashed line, separates the polyelectrolyte from the pure solution 
of salt.
The microscopic mobile ions --- 
counterions $(\oplus)$, cations $(+)$ and anions $(-)$ --- 
are free to move in the region $|\rb|>a$, and can cross the semipermeable 
membrane.}
\end{minipage}
\end{figure}

Our theoretical picture, exploiting the underlying periodicity
of the colloidal suspension \cite{Fu51}, consists of a spherical polyion
of radius $a$ with a uniform surface charge density, 
\be
 q_p(\rb)= -\fracd{Ze}{4\pi a^2} \delta(|\rb|-a)\:,
\ee
residing at the center of  a spherical {\it WS}
cell, where $Z$ is the polyion valence, $e$ is the electronic charge
and $\delta$ is the Dirac distribution.
Note that we assume that the polyion is negatively charged.
 The boundary of the
{\it WS}\/  cell corresponds to the semipermeable membrane.
The solvent, usually water, is modeled by a homogeneous 
continuum of a dielectric constant $\epsilon$. 
The mobile microions, modeled as point 
particles of charge $\pm e$, include the counterions and the ions 
derived from the dissociation of the monovalent salt, cations 
and anions. Inside the {\it WS} cell, these
are free to move within the annulus
$a<|\rb|<R$. The distribution of
microions is strongly determined by the electrostatic potential 
produced by the central polyion.  In the region $|\rb|>R$ the microions
 are assumed to be
unaffected by the presence of the polyion at $|\rb|=0$ and are uniformly
distributed \cite{text}.
Within the functional density theory, the microions inside the {\it WS}\/
cell are treated as an inhomogeneous fluid. At this level of approximation we
shall not distinguish between the counterions and  the cations of salt, 
so that
the  local density  of positive particles is
$\rho_+(\rb)$. The local density of co-ions
 (anions) is $\rho_-(\rb)$.

The {\it effective}\/ Helmholtz free energy inside the 
{\it WS}\/  cell, ${\cal F}^{\rm in}$,
associated with the region of the colloidal suspension, is composed
of:

\noindent 
(a) entropic ideal-gas contributions associated with the microion 
local densities $\rho_{\pm}(\rb)$,
\be
\beta F^{\rm ideal}_{\pm}= \int d^3\rb\,\rho_\pm(\rb) 
\left\{ \ln\left[\rho_\pm(\rb)\Lambda^3 \right] -1\right\}\:,
\ee
where $\beta=(k_B T)^{-1}$ is the inverse temperature
and $\Lambda=h/(2\pi mk_B T)^{1/2}$ is the thermal de Broglie wavelength;

\noindent
(b) electrostatic energy terms due to the polyion-microions interactions,
\be
F^{\rm el}_{p,\pm}=\pm  e \int  d^3\rb\,d^3\rb'\,
\frac{q_p(\rb)\rho_\pm(\rb')}{\epsilon |\rb-\rb'|}\:;
\ee

\noindent
(c) electrostatic energy terms due to the microions-microions 
interactions,
\be
F^{\rm el}_{ij}=
e^2 \int  d^3\rb\,d^3\rb'\,
\frac{\rho_i(\rb)\rho_j(\rb')}{\epsilon |\rb-\rb'|}\:,
\qquad \mbox{ for }i,j=\pm\:;
\ee

\noindent
(d)
and finally a (virtual) Lagrange-multiplier term,
\be
F^{D}=  \mu_D   \int d^3\rb\, \left[\rho_+(\rb) -\rho_-(\rb)\right]
\:, \label{eqn:lagrange}
\ee
which enforces the overall charge neutrality within the {\it WS}\/  cell, 
\be
\int\limits_{a<|\bms{r}|<R} 
d^3\rb\, \left[\rho_+(\rb) -\rho_-(\rb)\right] = Z \:. 
\label{eqn:constraint}
\ee
With the introduction of (\ref{eqn:lagrange}), 
we can treat the densities of microions, 
$\rho_\pm(\rb)$, as independent variables.
The Lagrange multiplier $\mu_D$ is traditionally referred to as 
the Donnan potential. After the minimization procedure is
complete, $\mu_D$ shall be eliminated 
using the charge-neutrality constraint (\ref{eqn:constraint}).
We have, then, the Helmholtz free energy inside the {\it WS}\/ 
cell,
\bea
{\cal F}^{\rm in}&=& F^{\rm ideal}_+ + F^{\rm ideal}_- + F^{\rm el}_{p,+} 
+ F^{\rm el}_{p,-} \nonumber\\
&&+ \frac12 F^{\rm el}_{++}
+ \frac12 F^{\rm el}_{--} - F^{\rm el}_{+-} + F^{D}\: .
\eea
   
In the region outside the {\it WS}\/  cell, the salt solution is, 
once again, charge neutral and the positive and negative 
ions are uniformly distributed. Due to this symmetry,
the overall Coulombic contribution vanishes up to the correlation
free energy, which we neglect within our  treatment \cite{text}. 
In the absence of an 
external field, the densities of positive and negative ions
within the electrolyte chamber are uniform and given by
\be
\rho_+(|\rb|>R)= \rho_-(|\rb|>R) = \rho_S \: ,
\ee
where $\rho_S$ is the (uniform) salt (cation/anion pairs)
concentration of the pure electrolyte in the chamber
after the equilibrium has been established.
The Helmholtz free energy, $F^{\rm out}$, then reduces to that of 
an ideal gas,
\be
\beta F^{\rm out}= V_S\,
 \rho_{+}\left[ \ln\left(\rho_{+}\Lambda^3 \right) -1 \right] +
 V_S\, \rho_{-}\left[ \ln\left(\rho_{-} \Lambda^3\right) -1 \right] \, ,
\ee
where $V_S$ is the volume of the chamber containing pure electrolyte.

Assuming that the region outside the {\it WS}\/  cell corresponds to the
chamber containing the pure-electrolyte solution, 
the system will achieve the Donnan equilibrium 
when the electrochemical potentials inside,
\end{multicols}
\ruleleft
\medskip

\be
\mu_{\pm}^{\rm in}(\rb)= \frac{\delta {\cal F}^{\rm in}}
{\delta \rho_{\pm}(\rb)} 
= \frac1{\beta}\ln\left[\rho_{\pm}(\rb)\Lambda^3 \right]\pm \mu_D \pm e \int  d^3\rb'\, \frac{q_p(\rb')+e\rho_+
(\rb')-e\rho_-(\rb')}{\epsilon |\rb-\rb'|}
= \frac1{\beta}\ln\left[\rho_{\pm}(\rb)\Lambda^3 \right]\pm \mu_D
\pm e\psi(\rb) \:,
\ee

\ruleright
\begin{multicols}{2}
\noindent{}and outside the {\it WS}\/  cell,
\be
\mu_{\pm}^{\rm out}=\frac1{V_S}
\frac{\partial F^{\rm out}}{\partial \rho_{\pm}} =  
\frac1{\beta}\ln\left(\rho_{\pm} \Lambda^3\right) = 
\frac1{\beta}\ln\left(\rho_{S} \Lambda^3\right) \:, 
\ee
are equal, 
\be
\mu_{\pm}^{\rm in}(\rb) = \mu_{\pm}^{\rm out} \, .
\ee
These conditions lead to the Boltzmann distribution of the
microion densities,
\be
\rho_{\pm}(\rb) =\rho_S \exp\left[
\mp\beta\mu_D\mp e\beta\psi(\rb)\right] \:.  \label{eqn:boltz}
\ee

On the other hand, the electrostatic potential $\psi(\rb)$ and 
the local charge densities obey the (exact) Poisson equation,
\be
-\nabla^2 \psi(\rb)=\nabla\cdot \Eb(\rb) = 
\frac{4\pi}{\epsilon}
 \left[ q_p(\rb)+e\rho_+ (\rb)-e\rho_-(\rb) \right] \:,  \label{eqn:poisson}
\ee
where $\Eb(\rb)$ is the electric field at the point $\rb$, which is
related to the electrostatic potential by $\Eb(\rb)=-\nabla \psi(\rb)$.
Inserting expression (\ref{eqn:boltz}) into Eq.~(\ref{eqn:poisson}) 
we find a Poisson-Boltzmann-{\it like}\/ equation 
\cite{Ma55,Al84,Gr91b,Ru81,Du92}. The effects of the correlations
inside the ionic atmosphere can be included through the use of the local
or the weighted-density contribution to the free-energy \cite{Gr91b}. 
This will be a topic for the future.

We now take advantage of the spherical symmetry of the system to eliminate
the angular dependence of the equations, that is, we  replace 
$\rb$ by $r=|\rb|$ in Eqs.~(\ref{eqn:boltz}) and (\ref{eqn:poisson}). 
The electric field also has a
spherical symmetry, so that $\Eb(\rb)=E(r)\fracd{\rb}{r}$.    
Integrating the Poisson equation~(\ref{eqn:poisson}) over a sphere of 
radius $r$ and using the divergence  theorem, we obtain the 
equation for the electric field strength $E(r)$, 
\bea
\int\limits_{|\bms{r}'|<r}d^3\rb'\,\nabla\cdot \Eb(\rb')
&=&\int\limits_{|\bms{r}'|=r}d\bm{S}'\cdot \Eb(\rb')
=4\pi r^2 E(r)\nonumber\\
&=& 
-\frac{4\pi Ze}{\epsilon} \left[1-\frac1Z \alpha(r)\right] \:,
\eea
where
\be
\alpha(r)=\int\limits_{|\bms{r}'|<r} d^3\rb'\,
\left[\rho_+(\rb')-\rho_-(\rb') \right]\:.
\ee
Using the Boltzmann equations~(\ref{eqn:boltz}) 
for the densities of microions inside the {\it WS}\/  
cell, we can then write
\be
\alpha(r) 
= 4\pi \rho_S \left[\exp\left(-\beta\mu_D\right)\alpha_+(r)
- \exp\left(\beta\mu_D\right)\alpha_-(r)\right] 
\label{eqn:alpha1} \:,
\ee
where $\alpha_{\pm}(r)$ are the radial integrals associated with 
the electrostatic potential $\psi(r)$, that can be written in terms 
of the electric field strength $E(r)$,
\bea
\alpha_{\pm}(r) &=& \int_{a}^r dr'\,r'^2 \, 
\exp \left[\mp e\beta \psi(r')\right]\nonumber\\
&=& \int_{a}^r dr'\,r'^2 \, 
\exp \left[\pm e\beta \int_a^{r'}dr''\,E(r'')\right] \:,
\eea
where we have chosen the gauge in which $\psi(r=a)=0$.
The Lagrange multiplier $\mu_D$ can now be eliminated using 
the charge-neutrality constraint over the 
{\it WS}\/  cell~(\ref{eqn:constraint}),
\end{multicols}
\ruleleft
\medskip

\be
\alpha(R)= 4\pi \rho_S \left[
 \exp\left(-\beta\mu_D\right)\alpha_+(R)  - \exp\left(\beta\mu_D\right)\alpha_-(R)\right] =Z \:,
\ee
which gives a quadratic equation for $\exp\left(-\beta\mu_D\right)$.
Solving the associated equation and replacing in (\ref{eqn:alpha1}), 
we obtain
\be
\alpha(r)= \frac{\sqrt{Z^2+(8\pi \rho_S)^2 
\alpha_+(R)\alpha_-(R)}}{2}
\left[\frac{\alpha_+(r)}{\alpha_+(R)}-\frac{\alpha_-(r)}{\alpha_-(R)}\right]
+ \frac{Z}{2} 
\left[\frac{\alpha_+(r)}{\alpha_+(R)}+\frac{\alpha_-(r)}{\alpha_-(R)}\right]
\:.
\label{eqn:alpha2}
\ee

\begin{multicols}{2}
To perform the numerical calculations, it is convenient to
write the equations in terms of dimensionless variables,
\be
\hat r=\frac{r}{\lambda}\,, \qquad
\hat E(\hat r) =e\beta\lambda E(r)\,,\label{eqn:alpha*}
\ee
\vspace{-1.6\baselineskip}

\bea
&&\hat\rho_\pm(\hat r)= 4\pi \lambda^3 \rho_\pm(r)\,,
\qquad \hat\rho_S =4\pi \lambda^3 \rho_S\,, \nonumber\\
\hat\alpha_{\pm}(\hat r)&=&\frac{1}{\lambda^3}\alpha_{\pm}(r)=
 \int_{\hat a}^{\hat r} d\hat r'\,\hat r'^2 \, 
\exp \left[\pm \int_{\hat a}^{\hat r'}d\hat r''\,\hat E(\hat r'')\right] \:,
\nonumber
\eea
where we have introduced the Bjerrum length, 
\be
\lambda=\frac{\beta e^2}{\epsilon}\:,
\ee
which measures the strength of the electrostatic interactions in the 
colloidal suspension. 

The equation for the dimensionless electric field strength can be written as
\be
\hat E(\hat r)=-\frac{Z}{\hat{r}^2}\left[1-\frac1Z \hat\alpha(\hat r)\right]
\:,
\label{eqn:field}
\ee
where $\hat\alpha(\hat r)$ is given by Eq.~(\ref{eqn:alpha2}),
replacing $\alpha_{\pm}(r)$ by $\hat\alpha_{\pm}(\hat r)$, 
and $8\pi\rho_S$ by $2\hat\rho_S$. Together with~(\ref{eqn:alpha2}), 
this is an integral equation which determines the reduced electric field
 $\hat E(\hat r)$. It  is gauge 
invariant, since it does not depend on the choice for $\psi(r=a)$, 
and already contains the boundary conditions,
\be
E(r=a)=-\frac{Ze}{\epsilon a^2}\:, \qquad\quad \mbox{and}\qquad\quad 
E(r=R)=0\:.
\ee 
 This equation can be solved numerically by replacing the 
continuous interval $\hat a<\hat r<\hat R$ by a finite grid and
iterating  (\ref{eqn:field}) until convergence is found.
In the absence of added salt, $\hat\rho_S=0$, we regain the 
results of previous calculations \cite{Gr91b,Al84}.

The dimensionless densities of microions inside the {\it WS}\/ 
cell are given by
\bea
\hat\rho_\pm(\hat r)&=&\frac{\sqrt{Z^2+4\hat\rho_S^2 
\hat\alpha_+(\hat R)\hat\alpha_-(\hat R)}\pm Z}{2\hat\alpha_\pm(\hat R)}\,
\times \nonumber\\
&& \times \, 
\exp \left[\pm\int_{\hat a}^{\hat r}d\hat r'\,\hat E(\hat r')\right] \:. 
\eea

The pressure inside the  cell is $P^{\rm in}=-\fracd{\partial 
\cal{F}^{\rm in}}{\partial V}=-\fracd{1}{4\pi R^2}
\fracd{\partial \cal{F}^{\rm in}}{\partial R}$, 
which after some algebra reduces to \cite{Ma55}
\be
\beta P^{\rm in}= \rho_+ (R)+\rho_-(R) \:. \label{eqn:pressurein}
\ee
The physical interpretation of~(\ref{eqn:pressurein}) is that at the 
border of the {\it WS}\/  cell the electrostatic forces on the
mobile ions vanish, since $E(R)=0$, and the pressure is given by the
ideal-gas law.
On the other side of the membrane the charge is 
uniformly distributed and the pressure,
up to the electrostatic correlations effects, is that of an
 ideal-gas,
\be
\beta P^{\rm out}= 
-\beta\frac{\partial{F}^{\rm out}}{\partial V_S}=2\rho_S \:.
\ee
Therefore, the osmotic pressure, defined as the difference between
the pressures inside and outside the {\it WS}\/  cell, is 
\bea
\beta \Delta P&=&\beta P^{\rm in}-\beta P^{\rm out}=
\rho_+ (R)+\rho_-(R) -2\rho_S \nonumber\\
&=& \frac{1}{4\pi\lambda^3}
\left[\hat\rho_+ (\hat R)+\hat\rho_-(\hat R) -2\hat\rho_S \right] \:.
\label{eqn:pressure}
\eea

\section{Comparison with experimental results}

The derivation in the previous section requires us to know
the concentration of salt $\rho_S$ once equilibrium across the membrane has 
been established. Although, in principle, this is possible to measure,
it is much simpler, from the point of view of experiment, if the 
osmotic pressure was calculated as a function of the concentrations 
of salt present in the two chambers, $n^0$ and $\rho_S^0$, 
{\it before}\/ the ion transport
is allowed to take place. Evidently it is these values that are directly 
controlled by the experimentalist. In this section we will present 
the results for the osmotic pressure as a function of the original 
concentration of salt added to the polyelectrolyte in the first chamber,
 $n^0$, and of the initial pure-electrolyte concentration 
 in the second chamber, $\rho_S^0$,
 before the ion transport is allowed to take place.
In this way our results can be directly compared with the measurements of
Benzing and Russel \cite{Be81}.

The initial concentration of salt in the pure-electrolyte
chamber, $\rho_S^0$, will not coincide with the same quantity
$\rho_S$ measured after the Donnan equilibrium is achieved. 
A relation between these two concentrations can, however, be
determined through a mass-conservation condition as follows. 
If $n^0$ and 
$n$ are the (macroscopic)
average salt concentrations in the colloidal-suspension 
chamber before and after the Donnan equilibrium is achieved, and
$V_C$ and $V_S$ are the volumes of the colloidal-suspension 
and the pure-electrolyte chambers, then the mass-conservation condition,
\be
n^0 V_C+ \rho_S^0  V_S=
n V_C  +  \rho_S V_S \:,
\ee
or 
\be
n = 
n^0(1+\gamma_1\gamma_2)-\gamma_1\rho_S\:,
\label{eqn:nsalt1}
\ee
gives a relation between the average concentrations of salt in the
suspension after and before the equilibrium in terms of the ratios
\be
\gamma_1=\frac{V_S}{V_C}\qquad\quad
\mbox{ and }\qquad\quad
\gamma_2=\frac{\rho_S^0}{n^0}\:.
\ee

Since the microions are just allowed to occupy an effective 
volume of the {\it WS}\/  cell, 
\be
V_{\rm eff} = \frac{4\pi}{3} (R^3-a^3)=
 \frac{4\pi\lambda^3}{3} \hat R^3(1-\phi)
\label{eqn:veff} \:,
\ee
where $\phi$ is the volume fraction occupied by the 
polyion, 
\be
\phi= \left(\frac{a}{R}\right)^3 \:,
\ee
it is useful to define an effective
salt concentration in the suspension, $n_S$,
given by    
\be
n_S =\frac{n }{1-\phi}=\frac{n^0(1+\gamma_1\gamma_2)-\gamma_1\rho_S}{1-\phi} \:.
\label{eqn:nsalt2}
\ee
This effective salt concentration is also 
obtained by integrating the density of negative salt particles
over the {\it WS}\/  cell,   
\be
n_S = \bar\rho_- = \frac12 \left(\bar\rho_+ + \bar\rho_- -
\frac{Z}{V_{\rm eff}}\right)\:, \label{eqn:ns}
\ee
where
\be
\bar\rho_{\pm} = \frac1{V_{\rm eff}} \int\limits_{a<|\bms{r}|<R}
 d^3\rb \, \rho_\pm(\rb)\:.
\ee

Using the charge-neutrality constraint over the {\it WS}\/  
cell~(\ref{eqn:constraint}),
we have
\be
\bar\rho_+ -\bar\rho_- = \frac{Z}{V_{\rm eff}} \:.
\label{eqn:constraint2}
\ee
On the other hand, integrating the Boltzmann equations~(\ref{eqn:boltz}),
we obtain
\bea
\bar\rho_\pm&=& \frac{4\pi\rho_S}{V_{\rm eff}}\exp(\mp\beta\mu_D)
\int_{a}^R dr'\,r'^2 \, 
\exp \left[\mp e\beta \psi(r')\right]\nonumber\\
&=& 
\frac{4\pi\rho_S}{V_{\rm eff}}\exp(\mp\beta\mu_D)\,
\alpha_{\pm}(R) \:,
\eea
which leads to the product
\be
\bar\rho_+\bar\rho_-=
\left(\frac{4\pi\rho_S}{V_{\rm eff}}\right)^2
\alpha_+(R)\alpha_-(R) \:. \label{eqn:rhoprod}
\ee
Solving~(\ref{eqn:constraint2}) and (\ref{eqn:rhoprod}), we find
\be
\bar\rho_\pm =\frac{Z}{2 V_{\rm eff}} 
\left\{\left[1+\left(\frac{8\pi\rho_S}{Z}\right)^2
\alpha_+(R)\alpha_-(R)
\right]^{1/2}\pm1\right\} \:.
\ee
Using now Eq.~(\ref{eqn:ns}),
we can relate the density of salt in
the pure-electrolyte chamber, $\rho_S$, and the effective concentration 
of salt in the colloidal suspension, $n_S$. 
After the equilibrium  is achieved,
 \be
(4\pi \rho_S)^2\alpha_+(R)\alpha_-(R)=
n_S^2 V_{\rm eff}^2 +Zn_ S V_{\rm eff} \:.
\label{eqn:rhosalt}
\ee
Substituting Eq.~(\ref{eqn:nsalt2}) for $n_S$,
we are left with a quadratic
equation  for $\rho_S$, which can be easily solved to yield
\end{multicols}
\ruleleft
\medskip

\be
\hat\rho_S=
\fracd{\frac{1}{3}\gamma_1\hat R^3 (Z+2\nu)-
\sqrt{\frac19\gamma_1^2\hat R^6Z^2+4\nu(Z+\nu)
\hat\alpha_+(\hat R)\hat\alpha_-(\hat R)}}
{2\left[\frac{\gamma_1^2}{9}\hat R^6-\hat\alpha_+
(\hat R)\hat\alpha_-(\hat R)\right]} \: , \label{eqn:rhofinal}
\ee
\vspace{-0.5\baselineskip}

\ruleright
\vspace{-0.5\baselineskip}
\begin{multicols}{2}
\noindent{}where
\be
\nu=n_S V_{\rm eff}+ \frac{\gamma_1}{3}\hat\rho_S\hat R^3  
=\frac{4\pi\lambda^3}{3} \hat R^3 n^0(1+\gamma_1\gamma_2) \:.
\ee

Inserting $\hat\rho_S$ into the expression for $\alpha(r)$,
the integral Eq.~(\ref{eqn:field}) can, once again, be solved 
iteratively to obtain the electric field and the density profiles
for the microions inside the {\it WS} cell. The osmotic pressure
is calculated as in the previous section,
but now is function of the initial densities
of salt $n^0$ and $\rho_S^0$. 

We are now ready to  compare our results for the osmotic pressure
 with the experimental measurements of
Benzing and Russel for monodisperse ordered 
polystyrene lattices \cite{Be81}.
In their experiments $V_C=V_S$ and the initial density of salt
inside the colloidal suspension was adjusted so that $n^0
=\rho_S^0 (1-\phi)$ to minimize the ion transport. We then have
\be
\gamma_1=1,\qquad\quad\mbox{ and }\qquad\quad \gamma_2=\frac{1}{1-\phi} \:.
\label{eqn:ratios}
\ee
The osmotic pressure was measured in terms of the volume fraction
$\phi$ for several initial salt concentrations 
$n^0$ for two distinct sizes of latex polyions.
The experimental results are reproduced in Figs.~2--3. From 
Eqs.~(\ref{eqn:pressure}), (\ref{eqn:rhofinal}) and 
(\ref{eqn:ratios}), we obtain the osmotic pressure for a fixed $Z$
in terms of the volume fraction $\phi$ for different initial
concentrations of salt $n^0$ inside the colloidal suspension.
In order to compare  the results of our theoretical
calculations with the measurements of Benzing and Russel \cite{Be81},
we need the valence $Z$ of the polyions.
We shall use $Z$ as a fitting parameter, determined in such a way
as to make the theoretical curve for the lowest electrolyte density
 passes through {\it one}\/ 
experimental data point.
 For instance,
for the latex A ($a=525$ \AA), we chose the experimental point
$n^0=10^{-7}$ mole/dm$^3$ 
and $\phi=0.3$, and we found 
for the polyion valence the value $Z_A=923$. 
Using always the
same value 
of $Z$ and varying $n^0$,
we obtained the other graphs (see Fig.~2). The same procedure was 
applied to the latex B ($a=1830$ \AA) --- see Fig.~3 --- 
where we have used the point 
$n^0=10^{-7}$ mole/dm$^3$
and $\phi=0.2$ and found  $Z_B=2212$. The general trend, observed both
in the experimental measurements as well as in the theoretical
calculations, is that the 
osmotic pressure was decreased by an increase in the
particle size or the amount of added salt.
 Comparing our theoretical
predictions with the experiments,
one might note that although the curves for
the lowest salt density 
provide an almost perfect fit to the experiment, as the electrolyte 
density is varied, such  a good agreement is lost. One possible
reason for this behavior might lie in the surface chemistry of
the latex particles, which causes the number of ionized sites $Z$ to be
dependent on the strength of electrolyte in  such a way as to make $Z$ 
{\it increase}\/ with the added concentration of salt \cite{Karin}.  
  
\section{Concluding remarks}

We have considered a system composed of a monodisperse 
charge-stabilized colloidal suspension and monovalent salt
separated from the pure electrolyte by a membrane
permeable to the microions and solvent, but 
impermeable to the polyions.
 Introducing a spherical {\it WS}\/  cell,
we have studied the crystalline phase within the framework
of the density functional theory. We have neither considered 
correlations between the microscopic free ions, nor the 
polyion-polyion interaction explicitly. 
The latter was replaced by an appropriate boundary
condition, namely, the charge-neutrality constraint over 
the {\it WS}\/  cell.

In particular, we have determined the osmotic pressure between the
two sides of the membrane as a function of the initial concentration
of added salt and the size, charge, and concentration of the 
colloidal particles. Using the valence of the polyion $Z$ as a
fitting parameter, we compared our theoretical results with 
experimental measurements for ordered polystyrene lattices  
of two different sizes \cite{Be81}, over a range of ionic 
strengths and colloidal volume fractions. 
Our results are in agreement with the experimental data, 
which shows that the osmotic pressure is 
decreased by an increase in the particle size or the amount 
of added salt, or by dilution of the colloidal particles in
the suspension. 
Although the experimental
data corresponding to the lowest salt concentrations 
 agrees reasonable well with our theoretical predictions, 
such good agreement is lost for higher electrolyte densities.
However, the trend of our results agrees with the 
fact that the valence $Z$ {\it increases}\/ with the salt 
concentration $\rho_S$ \cite{Karin}, 
which was not taken into account in the calculations.

\begin{figure}[ht]
\begin{center}
\leavevmode
\epsfxsize=0.45\textwidth
\epsfbox[25 20 580 450]{"
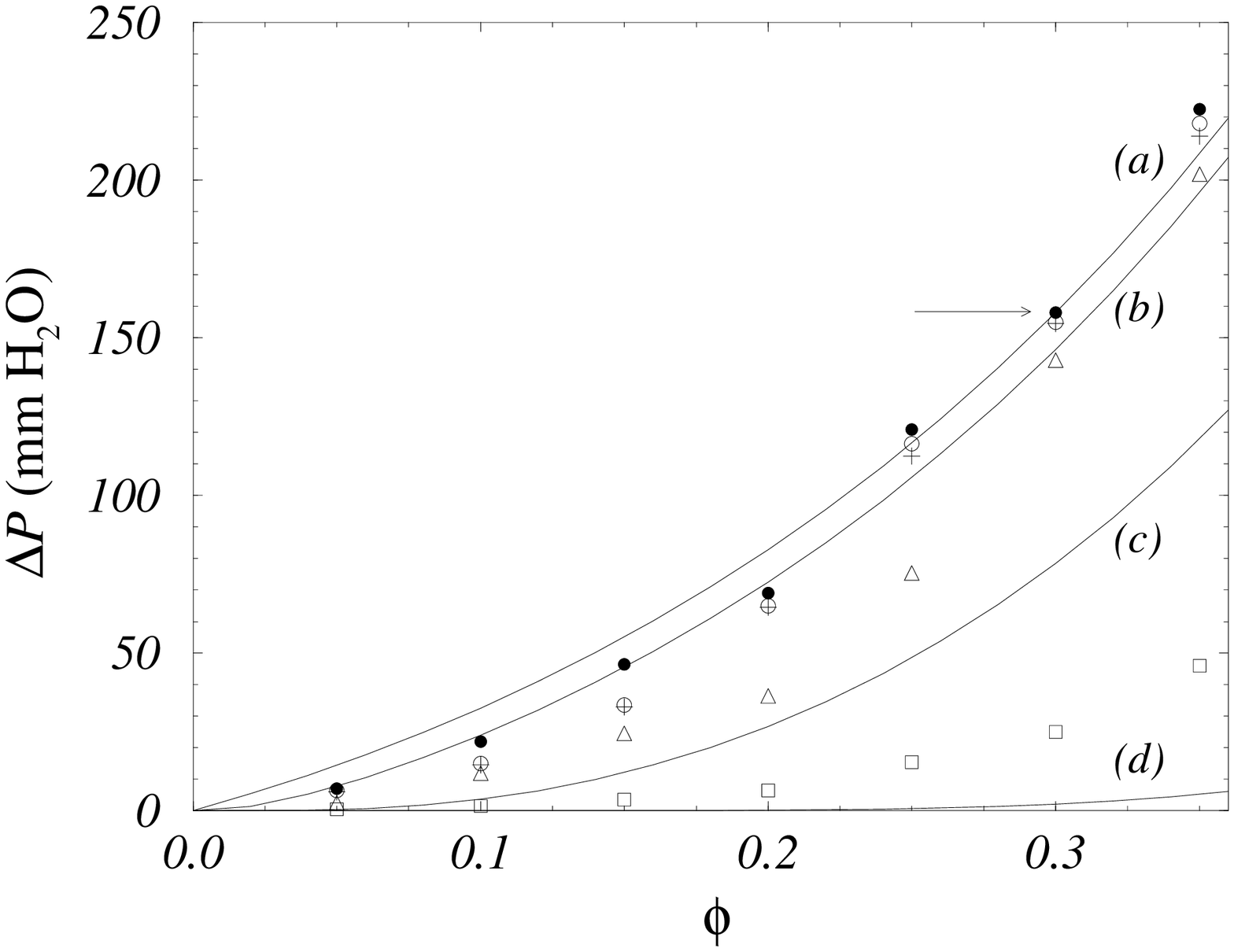"}
\end{center}
\begin{minipage}{0.48\textwidth}
\caption{Osmotic pressure $\Delta P$ versus volume fraction $\phi$
for latex A ($a=525$ \AA).
The symbols indicate experimental measurements from 
Benzing and Russel \protect\cite{Be81}, for several
initial concentrations $n^0$ of added electrolyte (NaCl) 
in the suspension (mole/dm$^3$): {\Large$\bullet$}, 0;
{\Large$\circ$}, 10$^{-6}$; $+$, 10$^{-5}$; $\triangle$, 10$^{-4}$; 
$\Box$, 10$^{-3}$. The arrow indicates the experimental point used to 
find the value of $Z_A=923$. The solid curves represent our corresponding 
theoretical predictions for different values of initial concentration
 $n^0$ of added salt in the suspension (mole/dm$^3$):
(a) 10$^{-7}$; (b) 10$^{-5}$; (c) 10$^{-4}$; (d) 10$^{-3}$. We have not
drawn the theoretical curve for $n^0=10^{-6}$
 mole/dm$^3$, since at this scale it is indistinguishable from the curve
for $n^0=10^{-7}$ mole/dm$^3$.}
\end{minipage}
\begin{center}
\leavevmode
\epsfxsize=0.45\textwidth
\epsfbox[25 20 580 450]{"
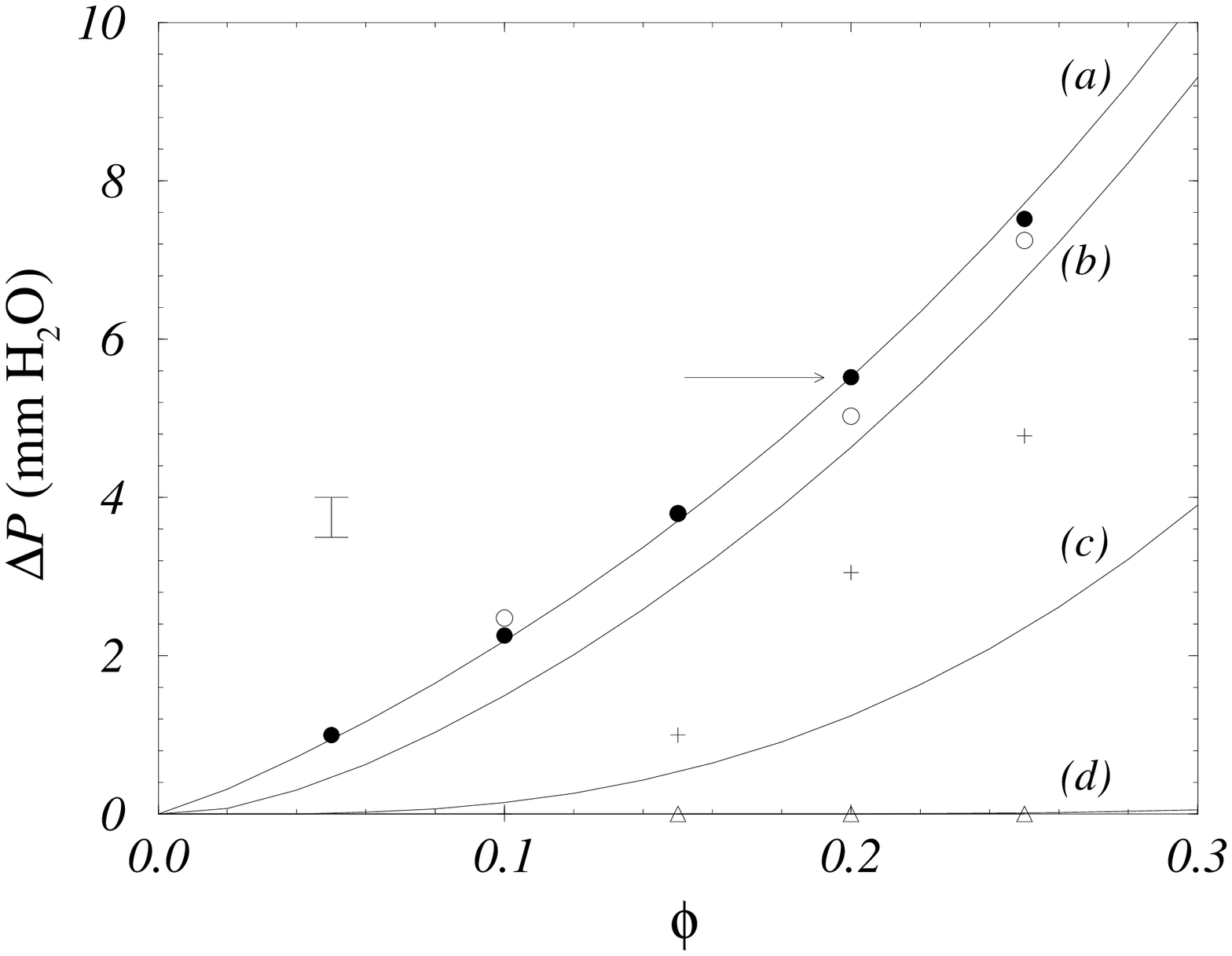"}
\end{center}
\begin{minipage}{0.48\textwidth}
\caption{Osmotic pressure $\Delta P$ versus volume fraction $\phi$
 for latex B ($a=1830$ \AA).
 The symbols indicate experimental measurements
from Benzing and Russel \protect\cite{Be81}, for several
initial concentrations $n^0$ of added electrolyte in the suspension, 
as identified in Fig.~2. The bar 
indicates the uncertainty in the experimental measurements.
The arrow indicates the experimental point used to find the 
value of $Z_B=2212$.
The solid curves represent our corresponding theoretical 
predictions for different values of initial concentration 
 $n^0$ of added
salt in the suspension (mole/dm$^3$):
(a) 10$^{-7}$; (b) 10$^{-6}$; (c) 10$^{-5}$; (d) 10$^{-4}$.}
\end{minipage}
\end{figure}

After the completion of this work, we became aware of the
experimental measurements, performed by Reus 
{\it et al.} \cite{Re97}, for a monodisperse 
charge-stabilized aqueous suspension of bromopolystyrene
particles.
They also compared their measurements  with the theoretical 
predictions using the Poisson-Boltzmann equation in a spherical {\it WS}\/  
cell \cite{Re97}, which is a calculation similar to ours of Section~II,
 requiring 
the knowledge of the equilibrium concentration, $\rho_S$, in the chamber
containing pure electrolyte. 
Although this was explicitly provided by Reus {\it et al.,} our 
formulation of Section~III allows us to predict the osmotic 
pressure based on the {\it a priori} knowledge of the amount of salt that
was being added to the system before the equilibrium is established.  
This is by far more practical than to wait 
for the establishment of the equilibrium and
only then try to measure the salt concentrations in the two compartments 
and use this as an input into the Poisson-Boltzmann equation. Furthermore,  
Reus {\it et al.} provided the explicit
value of the bare charge $Z$ obtained using conductimetric  titration 
and pH measurements.
In Fig.~4 we compare our theoretical predictions with the experimental 
data of Ref. \cite{Re97}. A perfect fit is found without any need for 
adjustable parameter.

\begin{figure}[ht]
\begin{center}
\leavevmode
\epsfxsize=0.45\textwidth
\epsfbox[25 20 630 450]{"
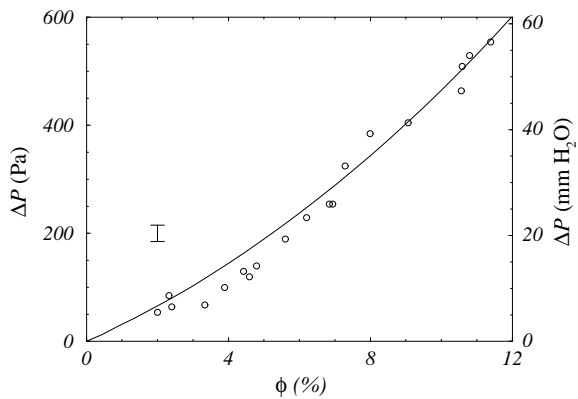"}
\end{center}
\begin{minipage}{0.48\textwidth}
\caption{Osmotic pressure $\Delta P$ versus volume fraction $\phi$
for a monodisperse charge-stabilized aqueous suspension of
bromopolystyrene particles of radius $a=510$ \AA.
The circles represent experimental measurements
from Reus {\it et al.} \protect\cite{Re97}, obtained for an equilibrium
concentration of salt in the pure electrolyte chamber 
$\rho_S=10^{-6}$ mole/dm$^3$. The bar 
indicates the uncertainty in the experimental measurements.
The solid curve is our theoretical result 
obtained using the experimentally determined valence of the
polyion, $Z=7500$. Thus, in this case, there are no fit  parameters.}
\end{minipage}
\end{figure}

Finally, we should add that Benzing and Russel \cite{Be81}
 have also tried to
estimate the value of the bare charge $Z$ using the electrophoretic 
mobility at infinite dilution. This can be directly related to the 
surface potential of
a colloidal particle by the Henry's law \cite{Be81}. Given the surface
 potential, the bare charge 
of the particle at infinite dilution can be obtained through the 
direct use of the Debye-H\"uckel theory. In Table~I the bare charges 
found through this procedure, $Z$(est), 
are compared with the ones we found by fitting the data, $Z$(fit). 
We see that the values obtained through the use of 
Henry's law fall way below the values needed to fit the data. In view 
of our perfect agreement with the data of  Reus {\it et al.} \cite{Re97},
 we are 
lead to the conclusion that there must be something wrong in trying 
to deduce the bare charge on the basis of electrophoretic 
mobility at infinite dilution. 

\begin{center}
\begin{minipage}{0.48\textwidth}
\begin{table}
\caption{Comparison between the 
estimates for $Z$ given by Benzing and Russel \protect\cite{Be81}, 
$Z$(est), and our values obtained through fit, $Z$(fit).}
\medskip
\begin{tabular}{cccc}
Latex& $a$ (\AA) & $Z$(est) & $Z$(fit)\\ \tableline
A & $525$& $174$ &  $923$\\
B & $1830$& $510$ & $2212$
\end{tabular}
\end{table}
\end{minipage}
\end{center}
 
\section*{Acknowledgments}

M. N. T. acknowledges helpful suggestions from Carlos S. O. Yokoi 
and discussions with Karin A. Riske and Carla Goldman.
 This work has been supported by the Brazilian agency CNPq
 (Conselho Nacional de Desenvolvimento Cient\'{\i}fico e Tecnol\'ogico).

\end{multicols}


\begin{thebibliography}{99}
\bibitem{text}This is a physical requirement considering that, in reality,
 the two 
compartments are macroscopic in size, but it also follows within our
 theoretical
picture from the overall charge neutrality, and a  mean-field assumption
that the total charge inside  the {\it WS}\/ cell is zero and the 
fluctuations are
neglected. This in turn implies that the average
electric field in the region $|\rb|> R$ is zero and the
microions are uniformly distributed. Once again we would 
like to  emphasize that, for the low 
concentrations of salt which we consider here (less than 
$10^{-4}$ mole/dm$^3$),
the electrostatic correlations effects can be neglected.
\end{thebibliography}
\end{document}